\documentclass[12pt]{revtex4}
\usepackage{amsmath}
\usepackage{amsfonts}
\usepackage{amsbsy}
\usepackage{lscape} 
\usepackage{color}
\usepackage{graphicx,epsfig}
\usepackage[english]{babel}
\usepackage{latexsym}
\usepackage{amssymb}

\newcommand{\beq}{\begin{equation}}
\newcommand{\eeq}{\end{equation}}

\newcommand{\eq}{{\rm eq}}
\newcommand{\lgr}{\left\lgroup}
\newcommand{\rgr}{\right\rgroup}
\newcommand{\Mpl}{M_{\rm Pl}}

\newcommand{\Tsph}{T_{\rm sph}}
\newcommand{\Gsph}{\Gamma_{\rm sph}}

\newcommand{\meff}{m^{\rm eff}_\nu}

\newcommand{\p}{\partial}
\newcommand{\mc}[1]{\mathcal{#1}}
\newcommand{\md}{\mathcal{D}}

\newcommand{\ov}{\overline}

\newcommand{\GeV}{{\rm GeV}}
\newcommand{\eV}{{\rm eV}}

\def\slashed#1{\setbox0=\hbox{$#1$}             
   \dimen0=\wd0                                 
   \setbox1=\hbox{/} \dimen1=\wd1               
   \ifdim\dimen0>\dimen1                        
      \rlap{\hbox to \dimen0{\hfil/\hfil}}      
      #1                                        
   \else                                        
      \rlap{\hbox to \dimen1{\hfil$#1$\hfil}}   
      /                                         
   \fi}                                        %

\begin{document}

\begin{titlepage}
\renewcommand{\thefootnote}{\fnsymbol{footnote}}

\setcounter{page}{1}

\vspace*{0.2in}

\begin{center}

\hspace*{-0.6cm}\parbox{17.5cm}{\Large \bf \begin{center}
$CPT$-odd Leptogenesis\end{center}}

\vspace*{0.5cm}
\normalsize

{\bf Pavel A. Bolokhov$^{\,(a,b)}$ and Maxim Pospelov$^{\,(a,c)}$ }

\smallskip
\medskip

$^{\,(a)}${\it Department of Physics and Astronomy, University of Victoria, \\
     Victoria, BC, V8P 1A1 Canada} \\
$^{\,(b)}${\it Theoretical Physics Department, St.Petersburg State University, Ulyanovskaya 1,
        Peterhof, St.Petersburg, 198504, Russia}\\
$^{\,(c)}${\it Perimeter Institute for Theoretical Physics, Waterloo,
ON, N2J 2W9, Canada}

\smallskip
\end{center}
\vskip0.2in

\centerline{\large\bf Abstract}

We calculate the baryon asymmetry of the Universe resulting from 
the combination of higher-dimensional Lorentz-noninvariant $CPT$-odd operators and 
dimension five operators that induce the majorana mass for neutrinos.
The strength of $CPT$-violating dimension five operators
capable of producing the observed value of baryon abundance is directly 
related to neutrino masses and found to be in 
the trans-Planckian range $(10^{-24}-10^{-22})~{\rm GeV}^{-1}$. Confronting it 
with observational tests of 
Lorentz symmetry, we find that this range of  Lorentz/$CPT$ violation 
 is strongly disfavored by the combination of the low-energy constraints and astrophysical 
data.

\vfil
\leftline{October 2006}

\end{titlepage}

%
%
\section{Introduction}
\label{intro}

Since the seminal paper by Sakharov \cite{Sakharov:1967dj}, it is well known that the 
baryon asymmetry of the Universe (BAU) can be generated dynamically, through the 
combination of baryon number violating processes, $C$ and $CP$ violation, and the 
departure from thermal equilibrium. It turns out that the Standard Model (SM) has all necessary  
ingredients for this to happen. Notably, the $B+L$ number is violated by the 
high-temperature sphaleron processes \cite{Klinkhamer:1984di},\cite{Kuzmin:1985mm}. However, the 
existing amount of $CP$-violation combined with tight constraints on the Higgs sector,
prevent efficient baryogenesis in the SM. Thus, BAU presents a formidable hint on physics 
beyond SM, and motivates new experimental searches for the extended 
electroweak sector and new sources of $CP$ violation.

	It is also known for some time that $CPT$-odd perturbations can effectively
	replace two Sakharov's conditions for baryogenesis: violation of $CP$
	invariance and the deviation from thermal equilibrium \cite{Dolgov:1981hv}.
	Indeed, a $CPT$-odd shift in the "mass" of a SM fermion (e.g. top quark \cite{Dolgov:1991fr}), 
    $\Delta m_{CPT}$ 
    would serve as an effective chemical shift between baryons and antibaryons above the 
    scale of the electroweak phase transition. 
    It is easy to see that $\Delta m_{CPT}/m_t\sim O(10^{-6})$ effect for top quark would be required to generate the 
    observed asymmetry  \cite{Dolgov:1991fr}. Unfortunately, at the level of the Lagrangian 
    is impossible to define a consistent 
    "$CPT$-odd mass" without breaking the Lorentz invariance.  $CPT$-odd mass would have to be identified 
with dimension three Lorentz-noninvariant operators \cite{Kost1}. Given the strength of
	constraints on lower-dimensional $CPT$/Lorentz noninvariant operators \cite{Colladay:1998fq}, one 
	has to conclude that lower dimensional operators 
    cannot be a source of the observed baryon asymmetry.

	The problem of $CPT$-odd baryogenesis was readdressed in \cite{Bertolami:1996cq}
and recently in \cite{Carroll:2005dj,Debnath:2005wk}. 
In \cite{Bertolami:1996cq} and \cite{Carroll:2005dj} among other options 
higher-dimensional $CPT$-odd operators 
were suggested as a source for baryon asymmetry. Suppose that a dimension five operator that 
shifts the dispersion relations of baryons relative to antibaryons is added on top of the SM. 
Let us further assume that initial value for $B-L$ is zero.
Then in the temperature range from $10^{10}$ to $10^2$ GeV where the sphaleron processes are in thermal 
equilibrium, the resulting baryon asymmetry will be determined by the amount of $CPT$ violation in the 
theory. If $CPT$-violating interactions are given by a dimension five operator parametrized by $1/\Lambda_{CPT}$, 
the inverse energy scale of $CPT$ violation, the resulting baryon 
asymmetry at the sphaleron freeze-out ($T\sim M_W$) will be given by 
\begin{equation}
\label{baryo}
	Y_b ~~=~~ \frac{\Delta b}{s} ~~\sim~~ \frac{T}{\Lambda_{CPT}}~ \sim \frac{M_W}{\Lambda_{CPT}},
\end{equation}
	where $ s $ is the entropy. It is clear then that $\Lambda_{CPT}< 10^{12}$ GeV will 
	be required to produce an observable asymmetry. 
Given the fact that both low-energy data and 
	astrophysical constraints limit a typical scale $\Lambda_{CPT}$ to be higher than the Planck scale,
	such scenario is completely ruled out.

    In this paper we explore the idea of the $CPT$-odd leptogenesis that is capable of 
    enhancing estimate (\ref{baryo}) by many orders of magnitude. 
    The main feature of any leptogenesis scenario is the use of the lepton number non-conservation 
    at high temperatures that results in a non-vanishing $ B - L $ 
    number, that is preserved by sphaleron processes \cite{Fukugita:1986hr}. 
	One of the advantages of leptogenesis is that the most natural way of mediating
the lepton-violating processes is through
	heavy majorana neutrinos, which also supply masses to the light neutrinos
	via the see-saw mechanism. 
	Heavy right-handed neutrinos with mass $ M_R $ mediate lepton number violating processes, 
	and thus keep lepton number violating processes in equilibrium 
until the temperature decreases to the point where the Hubble rate $ \Gamma_H $ begins to dominate
over the lepton-violation rate $ \Gamma_L $.
	In the assumption that Yukawa couplings are on the order one, this moment in Universe's 
	history can be determined as 
\[
	\Gamma_L ~~\propto~~ \frac{T^3}{M_R^2} ~~\sim~~ \Gamma_H ~~\propto~~ \frac{T^2}{\Mpl},
\]
	which gives an estimate for the temperature of the freeze-out for the $B-L$ number:
\[
	T_R ~~\propto~~ \frac{M_R^2}{\Mpl}~.
\]
	Therefore, in the scenarios of $CPT$-odd leptogenesis, one obtains the asymmetry
	which freezes out at $ T = T_R $ rather than at $ T = M_W $:
\[
	Y_{l(b)} ~~\sim~~ \frac{M_R^2}{\Mpl\Lambda_{CPT}}~.
\]
	Obviously, for 	$ M_R \sim 10^{15}~\GeV $ one gets a great enhancement 
	by $ T_R/M_W \sim 10^{9}$ over the $CPT$ baryogenesis scenarios  \eqref{baryo} where 
$B-L$ is zero.

    The purpose of this paper is to explore the $CPT$-odd leptogenesis scenario, 
    determine the required strength of the $CPT$-violating operators, and confront it with the 
    existing laboratory and astrophysics constraints. 
 For reasons explained earlier, we  concentrate on $CPT$-odd interactions of mass dimension five. 
	We introduce $CPT$-odd operators into the fermion sector of the Standard Model
\cite{MP:}:
\begin{equation}
\label{LV}
	\mathcal{L}_{LV} ~~=~~ \sum_{i=L,E,Q,U,D}\eta_i^{\mu\nu\rho}\cdot \ov{\psi}_i\gamma_\mu \md_\nu \md_\rho \psi_i~,
\end{equation}
	which cause an asymmetric shift of the dispersion relations for fermions and antifermions.
	Here $\eta_i^{\mu\nu\rho}$ is a symmetric irreducible Lorentz violating spurion field, that 
	can depend on the type of the SM fermions, and the summation extends over all 
	fields that carry the lepton or baryon number. The transmutation to the lower-dimensional 
	operators can be protected by the irreducibility condition, $\eta^{\mu\nu}_{~~\nu}=0$.
	A zeroth component of $ \eta_i^{\mu\nu\rho} $, $ \eta_i^{000} \equiv \eta_i $ in the 
reference frame where the primordial plasma is at rest provides 
	an asymmetric shift in the dispersion relations for particles and antiparticles.
    This way positive $\eta_{\rm lepton}$ creates a surplus of antileptons over leptons 
    in equilibrium which is maintained when the rate for the
lepton number violating processes is faster than the Hubble expansion.
	It is notable, that such $CPT$-odd perturbations allow for potential leptogenesis already
	with one flavor of heavy majorana neutrinos, whereas conventional leptogenesis requires
	at least two of them \cite{Fukugita:1986hr}.
	In the rest of this paper, we examine closely the kinetic equations for the 
    $L$($B$)-violating processes when $CPT$-odd shifts \eqref{LV}, lepton-number violation and 
    sphaleron processes are taken into account. 
	We adjust the coupling constants $ \eta $ in \eqref{LV} in such a way as to produce 
	the observed value of the baryon asymmetry and compare the results with the existing
	limits on Lorentz violating (LV) interactions. 
	We argue that the combination of bounds on LV from observations of high-energy cosmic rays
\cite{Gagnon:2004xh} and the low-energy clock comparison experiments
render the $CPT$-odd leptogenesis scenario fine-tuned  for models with operators of 
	mass dimension five \eqref{LV}, but allow it for higher-dimensional operators.

%
%
\section{Reaction rates and Boltzmann equations}
\label{rates}

	To demonstrate how the $CPT$-odd leptogenesis works we consider a model with only one heavy 
majorana neutrino. Its off-shell exchange mediates lepton number violating processes that 
	freeze out at the temperatures well below $ M_R $.
	At $ T > T_R $ these processes maintain the equilibrium value for the
	lepton number asymmetry. 
	In this section we calculate the rate of the lepton number violating processes and include it
	in the Boltzmann equations together with the sphaleron rate.
	
The mass term Lagrangian for heavy neutrinos reads as
\begin{equation}
	\mathcal{L}_m  ~~=~~ -\,\frac 12\, M_R\, \ov{N}{}_MN_M ~~+~~
				h_a\cdot \ov{L}_aHN_M ~~+~~  
				h_a^\dagger\cdot \ov{N}{}_MH^\dagger L_a~,
\end{equation}
	where $ N_M $ are singlet majorana neutrinos and $ h_a $ are the Yukawa couplings.
	We switch to Weyl spinors for convenience, in which the Lagrangian can be rewritten as
\begin{equation}
	\mathcal{L}_m  ~~=~~ 
	-\,\frac 12\, M_R\, \left( NN ~+~ \ov{N}\ov{N} \right) ~~+~~
				h_a\cdot \ov{L}_a\ov{ N}H ~~+~~  
				h_a^\dagger\cdot H^\dagger N L_a~,
\end{equation}
	where index $a$ runs over three different generations, and
\[	
	N_M ~~=~~ \left\lgroup 
		\begin{matrix}
			N_\alpha \\
			\ov{N}^{\dot{\alpha}}
		\end{matrix}
		\right\rgroup~.
\]
	Integrating out the heavy neutrinos, one obtains an effective lepton number violating vertex:
\begin{equation}
\label{L_eff}
	\mathcal{L}_{\rm eff} ~~=~~ \frac{Y_{ij}^\nu}{2\, M_R} \, H^\dag L_i^\alpha H^\dag L_{j\alpha}~
+{\rm h.c.},
\end{equation}
where $Y_{ab}^\nu=h_a^\dag h_b^\dag$.
	Substituting the vacuum expectation value 
for the Higgs field in \eqref{L_eff}
	creates a majorana mass term for light neutrinos. 
This interaction induces lepton number
violating processes which determine the lepton asymmetry
	until the lepton freeze-out. Alternatively, we could step by the stage with the heavy right-handed 
	neutrinos and postulate \eqref{L_eff} as a starting point in our analysis while taking $Y^\nu_{ab}$
	to be an arbitrary complex symmetric matrix.

Introduction of the $CPT$-odd interactions \eqref{LV} leads to the modification of 
dispersion relations for the SM leptons and antileptons. Taking lepton doublets, we neglect
the mass terms and find 
\begin{equation}
	E_L(p) ~~=~~ |\vec{p}| ~+~  \eta_L\, \vec{p}^2~, ~~~ 
E_{\bar L}(p) ~~=~~ |\vec{p}| ~-~  \eta_L\, \vec{p}^2~.
\label{modifiedE}
\end{equation}
Equation \eqref{modifiedE} leads to a shift in the equilibrium number density of leptons
\[
        n_L^\eq ~~=~~ \frac{g_L}{\pi^2 \beta^3}
			\left (\, 1 ~-~ \frac{12\,\eta_L}{\beta} \,\right)~,
\]
with the opposite sign of the shift for antileptons. Here $g_L$ is the total number of the spin, 
gauge and flavor
 degrees 
of freedom associated with electroweak doublets $L$, and 
$\beta$ is the inverse temperature. The difference,
\begin{equation}
\label{difference}
n_i^\eq - n_{\bar i}^\eq =-24\,\eta_i g_i(\pi^2 \beta^4)^{-1}, 
\end{equation}
where $ i = L $ for now,
represents an equilibrium lepton number 
induced by $CPT$ violation in the lepton doublet sector. 
As stated in the Introduction section, the final abundance can be roughly
estimated by evaluating the equilibrium density at the temperature
of the freeze-out.
A more accurate answer, however, can be obtained by analyzing Boltzmann equations in the 
presence of sphaleron processes and lepton number violation.

\begin{figure}
\includegraphics[width=9cm]{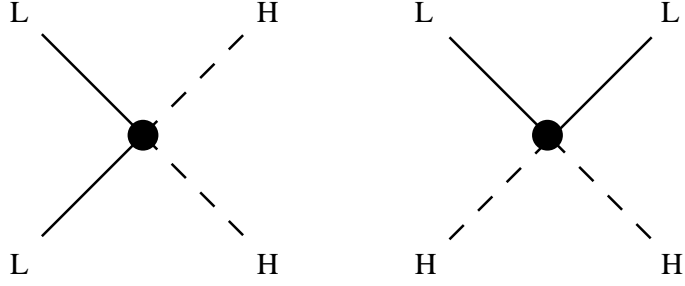}
\caption{$\Delta L=2$ processes generated by the effective vertex \eqref{L_eff}.}
\label{lflip_fig}
\end{figure}
There are two types of interactions induced by the effective
Lepton-Higgs vertex \cite{Luty:1992un,Plumacher:1996kc}, 
shown in  Fig.~\ref{lflip_fig}.
They generate the following processes relevant for leptogenesis:
\begin{align}
\notag
	L~~+~~L ~~\longleftrightarrow~~ H~~+~~H~  \\
\notag
	L~~+~~H ~~\longleftrightarrow~~ \ov{L}~~+~~H~,
\end{align}
with the same set of processes
for antileptons.
However, since the relevant part of the $CPT$-odd interactions is time reversal invariant, 
the amplitudes for
direct and inverse processes are equal, and we therefore have 
only three different amplitudes, which we call  
$ A_{LL} $, $ A_{\ov{LL}} $ and $ A_{LH} $.
Denoting the corresponding reaction rates (per unit volume) by
$ W_{LL} $, $ W_{\ov{LL}} $, $ W_{L{H}} $ and $ \ov{W}_{L{H}} $,
we have 
\begin{align*}
	W_{LL}   ~~=~~  &
		\int d\pi_p d\pi_q d\pi_k d\pi_r ~
		(2\pi)^4~ \delta^4 ( p + q - k - r )\,
		| A_{LL} |^2\, f_L^\eq(p)\, f_L^\eq(q)~, \\
	W_{\ov{LL}}   ~~=~~  &
		\int d\pi_p d\pi_q d\pi_k d\pi_r ~
		(2\pi)^4~ \delta^4 ( p + q - k - r )\,
		| A_{\ov{LL}} |^2\, f_{\ov{L}}^\eq(p)\, f_{\ov{L}}^\eq(q)~, \\
	W_{LH}  ~~=~~  &
		\int d\pi_p d\pi_q d\pi_k d\pi_r ~
		(2\pi)^4~ \delta^4 ( p + q - k - r )\,
		| A_{L{H}} |^2\, f_{{L}}^\eq(p)\, f_{{H}}^\eq(q)~, \\
	\ov{W}_{LH}  ~~=~~  &
		\int d\pi_p d\pi_q d\pi_k d\pi_r ~
		(2\pi)^4~ \delta^4 ( p + q - k - r )\,
		| A_{\ov{L}{H} }|^2\, f_{\ov{L}}^\eq(p)\, f_{H}^\eq(q)~,
\end{align*}
	where $ f_{L,H}^\eq(p) $ are the equilibrium distribution functions for Higgs fields
	and lepton doublets.
	In a toy model where only the lepton doublets and Higgs fields are present 
    one can immediately write the Boltzmann equations for the lepton number density as
\begin{equation}
\label{lepton_sys}
\begin{split}
	\lgr \p_t ~+~ 3\Gamma_H \rgr 
		n_L ~~=~~ &
	-\, 2\, W_{LL} \lgr \frac{n_L^2}{(n_L^\eq)^2} ~-~ 1 \rgr
	~~-~~
	W_{L\ov{H}} \lgr \frac{n_L}{n_L^\eq} ~-~ 
			\frac{n_{\ov{L}}}{n_{\ov{L}}^\eq} \rgr  \\
	\lgr \p_t ~+~ 3\Gamma_H \rgr 
		n_{\ov{L}} ~~=~~ &
	-\, 2\, W_{\ov{LL}} \lgr 
		\frac{n_{\ov{L}}^2}{(n_{\ov{L}}^\eq)^2} ~-~ 1 \rgr
	~~-~~
	\ov{W}_{L\ov{H}} \lgr \frac{n_{\ov{L}}}{n_{\ov{L}}^\eq} ~-~ 
			\frac{n_L}{n_L^\eq} \rgr.  \\
\end{split}
\end{equation}
	Here the Hubble rate is $\Gamma_H = 1.66 g_*^{1/2} T^2/\Mpl $ in terms of the 
total effective number of degrees of freedom $g_*$. 
The factor of two in the right hand side of \eqref{lepton_sys} reflects the fact 
	that the $LL$ processes change the number of leptons by two.
	An important thing to note is that even though we could have 
	modified the dispersion relations for the Higgs field, 
	its $CPT$-violating parameter would not enter the equations for the lepton number
	density at tree level.
	
	In order to generalize equations \eqref{lepton_sys} onto the full set of 
	SM fields, we introduce the effective parameters of $CPT$ violation in the 
	lepton and baryon sectors:
\begin{equation}
	\eta_l ~~=~~ \frac{ g_L \eta_L ~+~ g_E \eta_E } { g_L ~+~ g_E };~~~~~\eta_b ~~=~~ \frac{g_Q \eta_Q ~+~ g_U \eta_U ~+~ g_D \eta_D} 
			    {g_Q ~+~ g_U ~+~ g_D}~,
\label{effective}
  \end{equation}
where $g_i$ is the corresponding number of degrees of freedom in each sector. These parameters 
enter \eqref{difference} with $i=l,b$, and $g_l = g_L +g_E $,  $g_b= g_Q+g_U+g_D$.

	As already mentioned,
	one also has to include sphaleron processes, which affects 
	one linear combination of baryon and lepton number densities.
	The main effect of sphalerons is to wash-out $ B + L $,
	while keeping $ B - L $ intact. 
	Since the processes we consider occur far above the electroweak transition,  the sphaleron rate
has linear
	dependence on temperature 
\cite{Kuzmin:1985mm,Khlebnikov:1988sr}. 
	In the presence of $CPT$ violation, the sphaleron contribution to the Boltzmann equation
	for $ n_l $, $ n_b $ 
\cite{Kuzmin:1985mm,Moore:2000mx,Moore:2000ar} should be modified for the presence of the 
equilibrium baryon and lepton numbers \eqref{difference}:
\begin{equation}
\label{sph_corr}
	\p_t\, (n_b + n_l)
	~~~=~~~ -\, \Gsph 
		\lgr   n_b - n_b^\eq \;~+~\;
		       n_l - n_l^\eq 
		\rgr~,
\end{equation}
	where
\[
	\Gsph ~~\simeq~~ \omega\, T~, \qquad\qquad 
	{\rm ~with~}
	\omega ~\simeq~ \,10^{-5}~.
\]
	Equation \eqref{sph_corr} implies that $ B + L $ is washed out completely,
	and is somewhat simplifed relative to the realistic case.
	A detailed analysis shows (see {\em e.g.}
\cite{Harvey:1990qw})
	that the wash-out is only partial, with the final value of $B+L$ controlled by a nonzero $B-L$,
but we will employ the naive evolution equation
	\eqref{sph_corr}, arguing that the corrections to this equation are much smaller
	than the uncertainty with which $ \omega $ is known.
	
	Next we make a well-justified assumption of 
	smallness of the chemical potentials,
\[
	\frac{n_i}{n_i^\eq} ~~=~~ e^{\mu_i/T} ~~\simeq~~ 1 ~~+~~ \mu_i/T~,
\]
	which enables us to linearize the kinetic equations in $ \mu_i $.	
The kinetic equations for $ n_l$ take the following form:
\begin{equation}
\label{kinetic_eqn_prelim}
\begin{split}
	\lgr \p_t ~+~ 3\Gamma_H \rgr
		n_l ~~=~~ & 
	-\, \lgr 4\, W_{LL} ~+~ 2\, W_{L{H}} \rgr  \mu_l/T 
	~~-~~
	\omega\, T \lgr \mu_l/T ~+~ \mu_b/T \rgr 
	\\
	\lgr \p_t ~+~ 3\Gamma_H \rgr
		n_{\ov{l}} ~~=~~ &
	\phantom{-\,}
	\lgr 4\, W_{\ov{LL}} ~+~ 2\, \ov{W}_{L{H}} \rgr  \mu_l/T 
	~~+~~
	\omega\, T \lgr \mu_l/T ~+~ \mu_b/T \rgr ~.
\end{split}
\end{equation}

	For the (anti)baryons the kinetic equations are the same except
	that there are no contributions from the lepton number violating
	rates. 
A significant simplification comes from the smallness of the chemical potential.
	There are two possible sources for $CPT$-odd contributions 
	to the reaction rates in \eqref{lepton_sys}: modified dispersion relations and 
	$CPT$-odd modifications of thermal rates.  The smallness of $\mu_i/T$ allows us to neglect 
	any $CPT$-odd effects in the reaction rates in the right hand side of 
	\eqref{lepton_sys}, as
	they induce effects of the 2nd order in the $CPT$-violating parameter. Therefore, 
we take
	$ W_{\ov{LL}} = W_{LL} $ and
	$ \ov{W}_{L{H}} = W_{L{H}} $.

	From the above equations we only need their difference, the actual
	lepton (baryon) asymmetry.
	For convenience, we express the equilibrium number density in terms
	of the unmodified number density 
	$ n_i^0 = g_i/\pi^2 \cdot T^3 $
\[
	n_{i,\ov{i}}^\eq ~~=~~ n_i^0\, \left(\, 1 ~\mp~ 12\,\eta_i T \,\right)~,
	\qquad\qquad{i ~=~ l, b}~.
\]
	The asymmetries $ Y_i $ then can be defined as
\[
	n_i ~~-~~ n_{\ov{i}} ~~\equiv~~ 2\, n_i^0 \cdot Y_i~,
	\qquad\quad Y_i ~~=~~ \mu_i/T ~~-~~ 12\,\eta_i T~.
\]
	We also introduce a dimensionless parameter $ \gamma $, 
	by factoring out the dimensionful parameters $T^3/M_R^2$ from
	the rate of lepton number violating processes,
\begin{equation*}
	4\, W_{LL} ~~+~~ 2\, W_{L\ov{H}} ~~=~~ 
		 \gamma\, \frac{T^6}{M_R^2}~,
\end{equation*}
so that $\gamma$ scales as the fourth power of the neutrino Yukawa couplings or the 
sum of the squares of the eigenvalues of $Y_{ab}^\nu$: 
\begin{equation}
\label{gamma}
	\gamma  ~~=~~ \frac{3}{2\pi^2}\, \lgr \sum |h_a|^2 \rgr^2 = \frac{3}{2\pi^2}\, {\rm Tr}~ (Y^\nu_{\rm diag}
Y^{\nu\dagger}_{\rm diag})~.
\end{equation}

	Expressing equations \eqref{kinetic_eqn_prelim} in terms of
	$ Y_i $ and changing variables from time to temperature, we get:
\begin{equation}
\label{eqn_Y}
\begin{split}
	g_l \frac{d}{dT}Y_l 
	& ~~\;=~~\;
	\frac{0.6}{g_*^{1/2}}\, 
	\frac{\omega\,\Mpl}{T^2}\,
	\lgr g_l (\,Y_l~+~12\,\eta_l\,T\,) ~~+~~ 
	     g_b (\,Y_b~+~12\,\eta_b\,T\,)  \rgr 
	\;~~
	\\
	& \;~~+~~\;  
	\frac{0.6\,\pi^2}{g_*^{1/2}}\, 
	\frac{\gamma\,\Mpl}{M_R^2}\,
	\cdot (\, Y_l ~+~ 12\,\eta_l\,T\,)\\
	g_b \frac{d}{dT}Y_b 
	& ~~\;=~~\;
	\frac{0.6}{g_*^{1/2}}\, 
	\frac{\omega\,\Mpl}{T^2}\,
	\lgr g_l (\,Y_l~+~12\,\eta_l\,T\,) ~~+~~ 
	     g_b (\,Y_b~+~12\,\eta_b\,T\,)  \rgr .
\end{split}
\end{equation}
	The quantity of the ultimate interest is the baryon asymmetry 
	at the present time (normalized, e.g. on the photon number density,
	$ n_\gamma = s / 7.04 $ \cite{Kolb:1990vq}). 
	Using $ s = \frac{2\pi^2}{45} g_* T^3 $, one can express the experimentally measured 
	baryon to photon ratio via the asymmetry $Y_{b}$ that enters \eqref{eqn_Y},
\begin{equation}
\label{def_asy}
	\mathfrak{a}_B ~~=~~ 7.04\, \frac{45}{\pi^4}\, \frac{g_b}{g_*}\, Y_b
	~~\simeq~~ 0.6 \, Y_b 
		~~\equiv~~ (\, 6.1 ~\pm~ 0.3 \,)\times 10^{-10}~,
\end{equation}
where we  use $ g_b = 18 $ and $g_* = 106.75 $. 

	Note, that in the limit when the rate of sphaleron processes
	is very small, $ \Gsph \ll \Gamma_L $ ($ \Gamma_L $ is the rate
	of the lepton number violating processes), one can solve the kinetic equations
	exactly. 
	Taking $ \omega \to 0 $ in \eqref{eqn_Y}, we have:
\begin{equation}
\label{btz_simple}
	\frac{d}{dT}Y_l ~~=~~ \frac{\lambda\,\Mpl}{M_R^2} \,
			\lgr Y_l ~+~ 12\,\eta_l\, T \rgr~.
\end{equation}
	where we have introduced $
	\lambda ~~=~~ 0.6\, \pi^2
    		         (g_*^{1/2}\, g_l)^{-1}\,\gamma$. 
A solution that corresponds to $n_l$ close to equilibrium value at $T\gg T_R$ has the following form:
\begin{equation}
\label{btz_sol_simple}
	Y_l ~~=~~ -\,12\,\frac{\eta_l\,M_R^2}{\lambda\,\Mpl}
		~-~ 12\eta_l\,T ~,
\end{equation}
	which provides us with the expression for the lepton asymmetry:
\begin{equation}
\label{Y_L_simple}
	Y_l^{\rm fr} ~~=~~ -\, 12\,\frac{\eta_l\,M_R^2}{\lambda\,\Mpl}~.
\end{equation}
	The inclusion of sphalerons will diffuse approximately
	half of the lepton number yield into the baryon number
\cite{Kuzmin:1985mm}, so that Eq.~\eqref{Y_L_simple} is also an estimate for the BAU.

%
%
\section{The strength of $CPT$ violation derived from BAU}

In this section, we provide the numerical solutions to equations \eqref{eqn_Y},
determine the required strength of $CPT$ violation and confront it with 
existing experimental constraints.

To solve the system of kinetic equations, one has to add proper initial conditions.
	It is reasonable to impose these initial conditions at the temperatures
	where the essential part of leptogenesis begins, which we
	take to be $ M_R = 10^{15}~\GeV $:
	\begin{eqnarray}
	Y_l\bigl|_{M_R} & ~~=~~ Y_l^{\rm eq}, \\
	Y_b\bigl|_{M_R} & ~~=~~ 0~. \nonumber
\end{eqnarray}
	At high temperatures leptons and antileptons were in thermal and chemical equilibrium, which
	had a nonzero value of the lepton number defined by $ \eta_l $.
	This choice is quite sensible since 
	the freeze-out temperature $ T_R $ suggested by neutrino masses 
    is sufficiently smaller than $ M_R $.
	As for baryons, we impose symmetric $n_b=n_{\bar b}$ conditions 
	at high temperatures ($10^{15}$ GeV), as there are no fast processes that would bring 
$Y_b$ to the equilibrium position set by  $\eta_b$.

Since we chose to fix $M_R$, the only free parameters left are 
$\eta_i$ and $\eta_b$ parametrizing the strength of $CPT$ violation, 
and the neutrino Yukawa couplings. 
For the latter there is some natural range suggested by the 
oscillations among the light neutrino flavors. 
Introducing an ``effective'' neutrino mass that the kinetic equations 
\eqref{eqn_Y} depend on,
\begin{equation}
	\meff \equiv \left( \sum m_{\nu_i} \right)^{1/2}~~=~~ \left(\frac{\sum |Y^\nu_{\rm diag}|^2v^2}{2M_R}\right)^{1/2},
\end{equation}
we notice that $(\meff)^2$ is larger than any of the individual $\Delta m^2_{ij}$ measured 
in the oscillations experiments. Thus, taking the largest of $\Delta m^2_{ij}$ suggested by 
the oscillation of atmospheric neutrinos, 
$ \sqrt{\Delta m_{\rm atm}^2} \simeq 0.05~\eV $
\cite{Yao:2006px}
we find the following natural range for $\meff$:
\begin{equation}
\label{range_m}
	0.05~\eV\leq \meff\leq~~ 0.65~\eV~,
\end{equation}
where the upper limit comes from the cosmological
	bound on the sum of neutrino masses 
\cite{Hannestad:2003ye}.
	Defining the freeze-out temperature via relation
	$\Gamma_H(T_R)=\Gamma_L(T_R)$, one can translate 
	\eqref{range_m} to the realistic range of $T_R$:
		\begin{equation}
	10^{12}~\GeV ~~<~~ T_R ~~<~~ 10^{14}~\GeV~.
\end{equation}
		On the lower end of this range $ T_R $ overlaps with the 
	sphaleron ignition temperature $ \Tsph $, which is estimated
	to be of the order $ 10^{12}~\GeV $ \cite{Buchmuller:2005eh}.

The final result of our analysis is the prediction for the 
strength of $CPT$ violation in lepton and baryons sectors. 
Since equations \eqref{eqn_Y} are linear in $ Y_i $,
	it is sufficient to solve them numerically for two cases
\[
	\eta_l \neq 0,~~\eta_b = 0~~~~~ {\rm and}~~~~~ \eta_l = 0,~~\eta_b \neq 0~,
\]
	and then using the experimental value of BAU, fix the values of $ \eta_l $ and $ \eta_b $ 
	as  functions of $\meff$. 
	
	\begin{figure}
\includegraphics[width=9cm,angle=270]{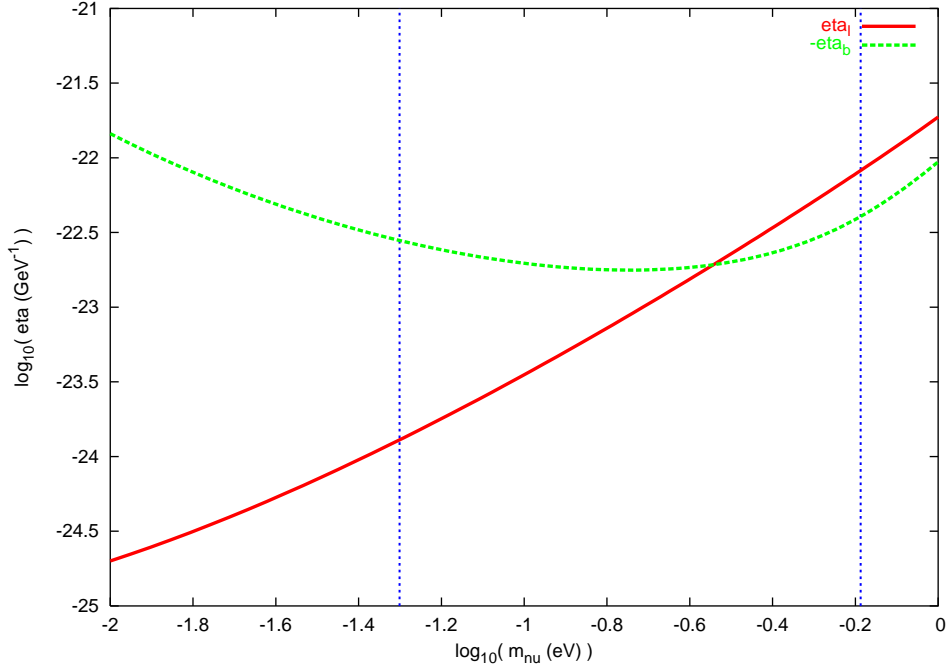}
\caption{$CPT$-odd parameters $ \eta_l $, $ \eta_b $ necessary to generate
	the observed BAU versus the effective neutrino mass.
	The left vertical line indicating the value of $ \meff $ suggested by the oscillation of atmospheric
neutrinos and the right vertical line showing the cosmological upper 
	limit on $ \meff $ \cite{Hannestad:2003ye}, bound the phenomenologically viable domain of $\meff$. }
\label{scan_fig}
\end{figure}
	
Fig.~\ref{scan_fig} exhibits the resulting dependence of $\eta_i$ on $\meff$ within a phenomenologically
	viable range of $ \meff $ bounded by two vertical dashed lines.
	One notices that $ \eta_b$ does not change much in the ``physical'' region.
	For $ \eta_l $-dominated scenario, in contrast, the increase of $\eta_l$ with 
	$ \meff $ is well pronounced. As expected, the lower mass $\meff$ leads to a 
{\em higher} freeze-out temperature $T_R$, and thus lower $\meff$ requires {\em lower}
$CPT$ violating parameter $\eta_l$ to get an observed value of BAU. Also not surprisingly, $\eta_l$ and 
$\eta_b$ required to reproduce BAU in our scenario have opposite signs. 
	
The lower end of the range \eqref{range_m} corresponds to a hierarchical 
scenario $m_1^2,m_2^2 \ll m_3^2$, with the tau-neutrino being the heaviest. 
The size of the $CPT$ violation suggested by the $CPT$-odd leptogenesis 
in this case is found to be
\begin{eqnarray}
\label{main_result}
	\eta_l = 9 \times 10^{-25} ~\GeV^{-1},~ \eta_b = 0
	~~~~~~~ {\rm or} ~~~~~~~ 
	\eta_l = 0, ~\eta_b = -1.5 \times 10^{-23} ~\GeV^{-1}~.
\end{eqnarray}
This is the main prediction of our work. 

\begin{figure}
\includegraphics[width=9cm,angle=270]{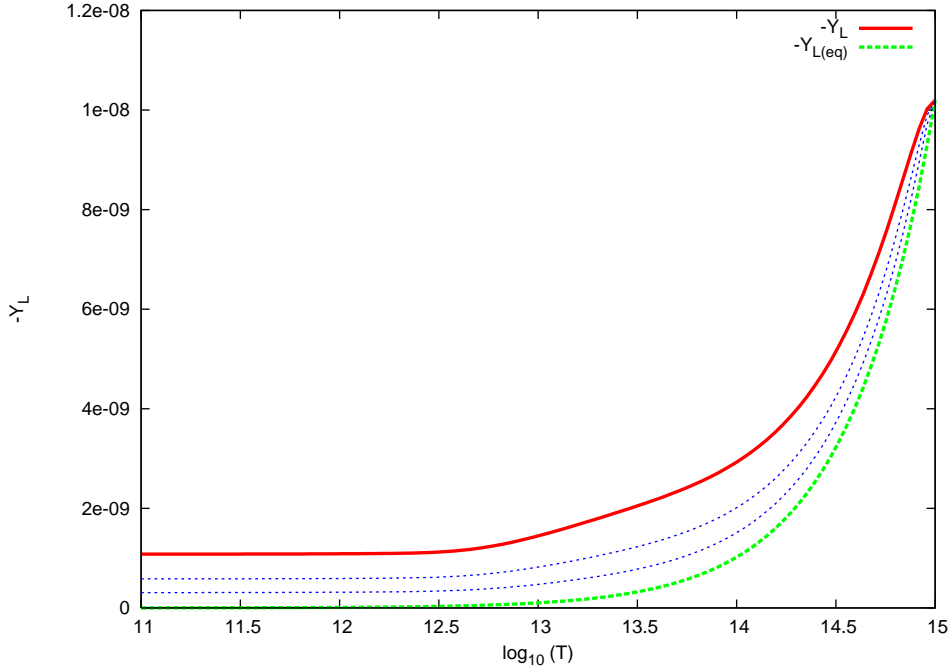}
\caption{
	Lepton asymmetry (solid line) and equilibrium lepton asymmetry (dashed line) 
driven by $CPT$ violation in the lepton sector
    for $\meff=0.05$ eV as function of temperature.
	The amount of $CPT$ violation is fixed to $ \eta_l = 9 \times 10^{-25}~\GeV^{-1} $ 
	to yield the observed value of baryon asymmetry.
	The final low-temperature plateau of $-Y_l$ equals to the baryon asymmetry $Y_b$. 
	The dotted lines correspond to $ \meff = 0.07~\eV $ and $ 0.10~\eV $, and 
	demonstrate the approach to the equilibrium curve with the increase 
	of mass $ \meff $.
	}
\label{l_dom_asymm_bau}
\end{figure}

\begin{figure}
\includegraphics[width=9cm,angle=270]{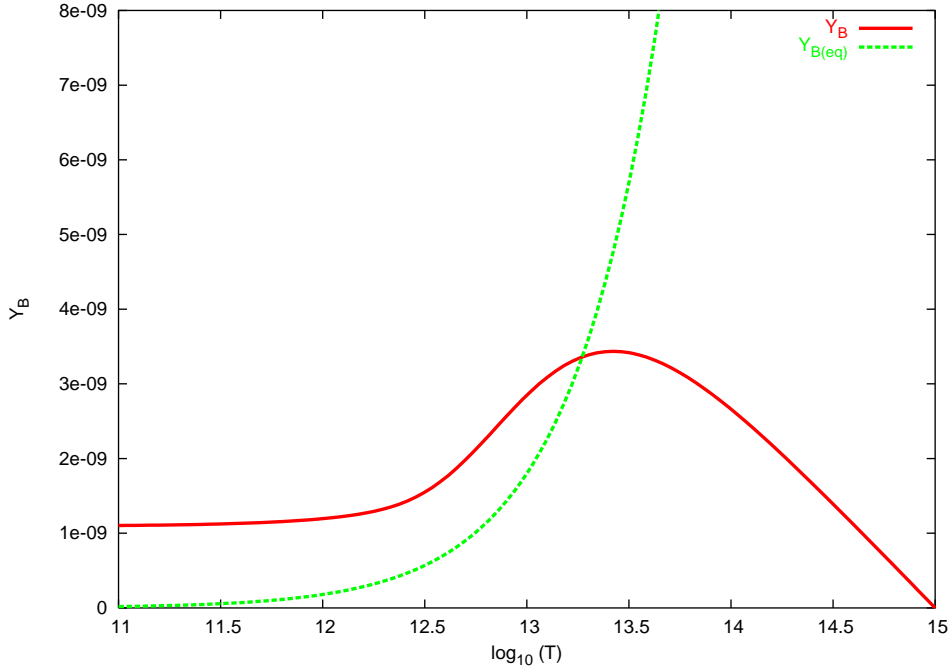}
\caption{Baryon asymmetry $ Y_b $ and equilibrium baryon asymmetry vs 
	temperature with $CPT$ violation concentrated in  the baryon sector. The parameters  
$\meff=0.05$ eV and $\eta_b = -1.5 \times 10^{-23}~\GeV^{-1}$ are chosen to match the observed asymmetry.}
\label{b_dom_asymm_bau}
\end{figure}

	Figures~\ref{l_dom_asymm_bau} and \ref{b_dom_asymm_bau} illustrate the 
	case of $\meff=0.05$ eV in more detail, by showing the evolution of the baryon/lepton
	asymmetry as a function of temperature. When $CPT$ violation is concentrated in the 
lepton sector, see Fig.~\ref{l_dom_asymm_bau}, the lepton asymmetry follows the 
equilibrium value of the (lepton)
	asymmetry at high temperatures to freeze out below $10^{14}$ GeV.
When $CPT$ violation is given by $\eta_b$, the asymmetry $ Y_b $ starts from zero, 
    overshoots the equilibrium curve just above $10^{13}$ GeV, to freeze out
	at lower temperatures.


\section{Comparison with experimental constraints on $CPT$ violation}

	Now we are ready to confront our predictions for $CPT$-violation (\ref{main_result}) 
with the experimental constraints on it. The modification of dispersion relation by 
dimension five operators has been discussed at length in the literature. Below we 
list a set of relevant constraints on dimension five operators in the 
fermionic sector of the SM and comment on their applicability:
\begin{align}
\notag
	|\eta_d-\eta_Q - 0.5(\eta_u-\eta_Q)| &~<~ 10^{-27}~{\rm GeV}^{-1}~, && \text{\cite{Sudarsky:2002ue,MP:}}\\
\notag
	|\eta_L|,~|\eta_E| &~<~ 10^{-20}~{\rm GeV}^{-1}~,    && \text{\cite{Jacobson:2005bg}}\\
\notag
	|\eta_L|, ~|\eta_E| &~<~ 10^{-33}~{\rm GeV}^{-1}~.  && \text{\cite{Gagnon:2004xh}}
\end{align}
	The first constraint arises because the axial-vector-like combinations of $\eta_i$ in the quark sector 
	lead to the coupling of the nucleon spin to a preferred direction. In models where the 
	preferred frame is associated with the rest frame of the cosmic microwave background, the net spin energy shift 
is proportional to the velocity of the lab frame $v\sim O(10^{-3})$, $\Lambda_{QCD}^2\eta_i(v\cdot s)$, 
which is to be compared with 
the experimental sensitivity $10^{-31}$ GeV \cite{clock1,clock2}. The low-energy constraints on lepton 
operators are considerably weaker \cite{MP:}.
The strongest constraints on dimension five 
$CPT$-odd operators in the lepton sector come from considerations of $p\to p l\bar l$ processes that become 
energetically allowed and prevent acceleration of protons to energies of $\sim 10^{21}$ eV. It is important 
that constraints \cite{Gagnon:2004xh} are double-sided, which is the consequence of asymmetric modification 
of dispersion relation for leptons and antileptons \eqref{modifiedE}. 

The strength of $CPT$ violation in the lepton sector 
derived from the baryon asymmetry \eqref{main_result} is consistent with the astrophysical bounds on 
$CPT$-violating QED \cite{Jacobson:2005bg}, but appears to be grossly inconsistent with \cite{Gagnon:2004xh}.
In fact, typical constraints on dimension five operators \cite{Gagnon:2004xh} derived from the existence of 
the high-energy cosmic rays appear to destroy any hopes for the $CPT$-odd baryogenesis, even if the 
scale of $T_R$ is pushed all the way up to the Planck scale.  It is easy to see, however, that this is 
not the case. If  $CPT$-odd sources in the quark sector dominate over the lepton sources by a factor of 
20-30, strong constraints on $CPT$ violation might be avoided. If, for example, among the 
$CPT$-odd sources the right-handed up quark has the largest modification of its dispersion relation, 
the energetically favored process $p\to \Delta^{++}\pi^-$ allows the ultra high-energy cosmic rays to exist in the 
form of $\Delta^{++}$, an option which cannot be observationally ruled out \cite{Gagnon:2004xh}. 
It is very important to observe that 
the {\em negative} sign of $\eta_U$ suggested by BAU \eqref{main_result} is exactly the sign of $\eta_U$ 
needed for $p\to \Delta^{++}\pi^-$ to happen at high energies.  
Nevertheless, the required size of $\eta_U$, $\eta_U \sim - (10^{-23}-10^{-22})~\GeV^{-1}$
appears to be in sharp conflict with the low-energy constraints \cite{Sudarsky:2002ue,MP:}, and at least 
four orders of magnitude tuning for dimension five sources is needed. 
This consideration shows an important complementarity between the
astrophysical bounds on Lorentz violation and low-energy searches of the breakdown of rotational 
invariance.

We can extend our analysis to theories where $CPT$ violation comes from operators of dimension 
seven, nine, etc., should for some contrived reasons lower dimensional operators be absent. 
	We note that to sufficient accuracy, the resulting BAU
	will be determined by the equilibrium lepton asymmetry
	at the freeze-out time, $ \eta^{(7)}\,T_R^3 $, $ \eta^{(9)}\,T_R^3 $, 
	where $\eta^{(n)}$ parametrize the strengths of the higher dimensional operators:
\[
	\mc{L} ~~=~~ \sum \eta^{(n)}_{\kappa\mu...\nu}\,
	\ov{\psi} \gamma^\kappa \md^\mu ...\md^\nu  \psi.
\]
As before, the transmutation to lower-dimensional operators can be forbidden 
by the irreducibility of $\eta^{(n)}$ tensors. 

The low-energy constraints on dimension seven and higher $CPT$-odd operators 
are totally irrelevant, as the possible influence on the nucleon spin is 
suppressed by extra power(s) of $(\Lambda_{QCD}/\Lambda_{CPT})^2$.  The constraints 
coming from the propagation of the high-energy cosmic rays are harder to avoid, as their 
relative strength scales down as $(E_{max}/\Lambda_{CPT})^2$, where $E_{max}$ is the maximal 
energy of the high-energy cosmic rays $E_{max}\sim 10^{12}$ GeV. In fact, since the decoupling
temperature $T_R$ can only be  marginally larger than $10^{12}$ GeV, the $CPT$-violating 
sources of dimension seven in the lepton sector allowed by the cosmic rays would not 
be able to produce the required size of the baryon asymmetry. However, the same loophole 
with the stability of $\Delta^{++}$ at high-energies 
exists for the dimension seven operators, and the right-handed up-quark $CPT$ violation
at the level of 
\begin{equation}
\eta^{(7)}_U = - [(10^{17}-10^{18})~ {\rm GeV}]^{-3}
\end{equation}
results in the right magnitude of BAU while avoiding all 
experimental constraints on Lorentz and $CPT$ violation.

%
%
\section{Discussion}

	We have seen that the presence of $CPT$-odd interactions is theoretically capable
	of replacing \emph{two} of Sakharov's conditions of baryogenesis: non-conservation 
	of $CP$ symmetry and departure from thermodynamical equilibrium.
	The reason for this is that non-zero lepton (or baryon) asymmetry can develop even in 
thermal equilibrium if  the $CPT$-violating shifts of dispersion relations for particles 
and antiparticles and fermion number violating processes are operative at the 
same time \cite{Dolgov:1981hv}. In this paper, we considered in detail the idea of 
leptogenesis driven by $CPT$-violating sources in the fermionic sector of the 
Standard Model. In this scenario, the generation of the $B-L$ number 
occurs at temperatures of about $10^{12}-10^{14}$ GeV, which results in a huge enhancement of the 
asymmetry as compared to the $CPT$-odd electroweak baryogenesis scenario, where $B-L=0$ and 
the equilibrium value for $B+L$ is maintained until the electroweak breaking, 
	$T\sim 100~\GeV$. 
Consequently, the $CPT$-odd leptogenesis requires only trans-Planckian size of 
$CPT$ violation, $\eta_i \sim 10^{-22}-10^{-24}~\GeV^{-1}$. 

We believe that this is the minimal level 
of $CPT$ violation required to reproduce the observable asymmetry. Lower levels of $CPT$-breaking 
may generate BAU 
only at the expense of  raising the decoupling temperature for $B-L$ processes, to the range of 
the {\em e.g.} GUT scale. Models with such a high initial temperature possess very serious cosmological 
problems of their own related to the overproduction of dangerous relics (monopoles, gravitinos), and 
are difficult to incorporate into inflation.

   The most natural models of $CP$-odd leptogenesis require two heavy neutrino singlets
   to work. We have shown that one species is perfectly sufficient for the $CPT$-odd scenario. 
   In fact, one could take even more conservative approach and associate the majorana 
   masses of light neutrinos with the effective Lorentz-conserving interaction \eqref{L_eff} without 
   specifying its origin. The $CPT$-odd leptogenesis in this case will proceed exactly as 
described in the paper, as long as \eqref{L_eff} remains unsuppressed at high energies. 
As a consequence of  the reduced heavy sector, the 
connection to the phenomenology of light neutrinos becomes more 
direct. As shown, the rate of the lepton-number violating processes is 
directly proportional to the sum of the mass squared of all light neutrino species. 
	
	Confronting the predicted size of $CPT$-violation with the existing experimental 
	and astrophysical constraints we find that both the low-energy precision searches of preferred 
	directions and the astrophysical constraints derived from the 
existence of charged high-energy cosmic rays puts severe 
	constraints on $CPT$-odd leptogenesis. The latter, being 
especially stringent, rules out a possibility of $CPT$-odd leptogenesis
driven by $\eta_l$ when $\eta_b =0 $. The inverse case, $\eta_l=0;~\eta_b \neq0 $
cannot be ruled out from the astrophysical considerations, as the bounds would not apply
if {\em e.g.} the $CPT$ violation is concentrated in the right-handed up-quark sector. 
In this case, however, one should expect  sizable effects in the clock comparison 
experiments. Current sensitivity to such operators is at the level of $10^{-27}~\GeV^{-1}$, 
and thus would require at least four orders of magnitude fine-tuning to make \eqref{main_result} 
evade the bounds.

	The $CPT$-odd interactions that modify dispersion relations 
	represent a relatively small subset of dimension five $CPT$-odd interactions \cite{BP}. 
	Is it feasible that other operators could  drive (baryo)leptogenesis while 
evading strong astrophysical and laboratory constraints? If physics responsible for $CPT$ violation 
preserves supersymmetry, operators that modify dispersion relations are simply not allowed 
\cite{GrootNibbelink:2004za,Bolokhov:2005cj}. Instead, a different class of $CPT$-odd operators may appear: 
\begin{equation}
\bar L \gamma_\mu L H^\dagger H,~~\bar Q \gamma_\mu Q H^\dagger H, ~~{\rm etc}. 
\label{alternative}
\end{equation}
When the lepton or 
baryon number is calculated in equilibrium, such operators will create an 
effective chemical potential that grows with temperature, $\mu \sim T^2\zeta$, where 
$\zeta$ parametrizes the strength of $CPT$ violation. The easiest way to see that is to 
consider the thermal field theory correlator between the baryon/lepton number density and 
such $CPT$-odd operators. Inside a thermal loop, the Higgs field bilinear will produce $T^2$,
and the scaling of the effective chemical potential with temperature will be exactly the 
same as in the case of $\eta_i$ operators. Although operators \eqref{alternative} do not influence 
the propagation of the high-energy cosmic rays, they have a phenomenological "problem" of their own. 
Inside loops such operators create quadratic divergencies and generate dimension three $CPT$-odd 
operators proportional to the square of the ultraviolet cutoff. In the most UV-protected case, 
the role of this cutoff is assumed by the supersymmetric soft-breaking scale. Still, the strength
of typical constraints is on the order of $10^{-10}M_{\rm Pl}^{-1}$ 
\cite{Bolokhov:2005cj}, making the 
scenario driven by \eqref{alternative} fine-tuned below 1 ppm level. 
Finally, what if $CPT$-violation is concentrated in the heavy right-handed neutrino sector? 
Phenomenology of such model was addressed in \cite{Mocioiu:2002pz}, where it was shown that 
loop effects reintroduce $CPT$ violation in the sector of charged leptons. 
Upon integrating out heavy neutrino 
fields, one produces operators similar to \eqref{alternative}, and therefore 
such possibility is also fine-tuned. 

Our main conclusion is that the natural levels of $CPT$/Lorentz violation 
suggested by the $CPT$-odd (lepto)baryogenesis scenario are $10^{-3}-10^{-5}$
in the Planck mass units, which is well within the ranges already disfavored by the laboratory experiments 
and observations of the 
high-energy cosmic rays. This analysis 
relies on the spurion approach to $CPT$ violation, which assumes that the strength of the 
$CPT$-odd source was essentially the same in the early Universe and today. 
It is of course conceivable that the dynamical effects could have been responsible for the 
$CPT$ breaking at high temperatures, sourcing the baryogenesis, with relaxation of $CPT$ sources to zero 
at the later stage \cite{Cohen:1987vi}.

{\bf Acknowledgements.} The authors would like to thank A. Ritz for useful discussions.
This work is supported in part by the NSERC of Canada. Research at the Perimeter Institute 
is supported in part by the Government
of Canada through NSERC and by the Province of Ontario through MEDT. MP would like to thank 
the Aspen Center for Physics, where part of this research project was carried out. 
PAB thanks the Pacific Institute for Mathematical Sciences for its support enabling the participation in the 2006 
Summer School on Strings, Gravity, and Cosmology.

\bibliographystyle{apsrev}
\bibliography{genesis}

\end{document}